\documentclass[8pt,preprint2]{aastex}

\usepackage{natbib}
\slugcomment{LA-UR-05-7544}
\shorttitle{H2 in white dwarf atmospheres}
\shortauthors{Kowalski}

\def\wig#1{\mathrel{\hbox{\hbox to 0pt{%
          \lower.5ex\hbox{$\sim$}\hss}\raise.4ex\hbox{$#1$}}}}

\begin{document}

\title{On the Dissociation Equilibrium of $\rm H_2$ in Very Cool, Helium-Rich White Dwarf Atmospheres}
\author{Piotr M. Kowalski 
}

\affil{Department of Physics and Astronomy, Vanderbilt University, Nashville, TN 
37235-1807}
\affil{Los Alamos National Laboratory, MS F699, Los Alamos, NM 87545}
\email{kowalski@lanl.gov}

\begin{abstract}

We investigate the dissociation equilibrium of $\rm H_2$ in very cool, helium-rich white dwarf atmospheres.
We present the solution of the non-ideal chemical equilibrium for the dissociation of molecular hydrogen in a medium of dense helium.
We find that at the photosphere of cool white dwarfs of $T_{\rm eff}\rm=4000 \, K$, the non-ideality results in an increase 
of the mole fraction of molecular hydrogen by up to a factor of $\sim 10$, compared to the equilibrium value for the ideal gas. 
This increases the $\rm H_{2}-He$ CIA opacity by an order of magnitude and will affect the determination of the abundance of hydrogen
in very cool, helium-rich white dwarfs.
\end{abstract}


\section{Introduction}

Several very cool white dwarfs with suspected $T_{\rm eff}\rm \wig< 4500 \, K$ have been discovered recently \citep{FAR,KM,G,OPP,Harris01,HOD,Ibata00,HR2}. 
Most of them are thought to posses helium-rich atmospheres with an very high $\rm He/ H \wig> 10^{3}$ ratio \citep{BAL,KM,G,BL,Bergeron01,OPP,HOD}.
In most cases, however, current atmosphere models fail to reproduce the observed spectra and photometry  
of these peculiar stars.  The reason, and there may be more than one, for this shortcoming of the models
is presently unknown. However, current models predict extreme physical atmospheric 
conditions for such stars, reaching densities of up to $\rm 2-3 \ g/cm^{3}$.  Under these conditions, the mostly ideal gas constitutive
physics used in published atmosphere models is demonstrably inadequate. A careful look at the dense matter
effects on the equation of state, chemistry, opacities, and radiative transfer is necessary to compute physically realistic models of these stars.
Several of these effects have been studied previously, such as refractive radiative transfer \citep{KS04}, the effects of fluid
correlations on $\rm He^-$ free-free and He Rayleigh scattering \citep{KSM05,IRS}, and the ionization of warm, dense helium \citep{KSM05,BSW}.
In this contribution we present an additional correction that arises in the dense fluid: The solution for the dissociation of molecular 
hydrogen in dense fluid helium, in the limit $\rm He/H>>1$.  
The relative importance of these corrections varies considerably, even more so when they are combined.
As several more dense matter effects remain unexplored, it is premature to ponder their implications for the analysis of the coolest
white dwarfs known and whether they will result in models that reproduce the data.  Nonetheless, incorporating adequate constitutive 
physics in atmosphere models is a necessary step to reach a proper understanding of these peculiar stars.

We introduce non-ideal effects into the equilibrium dissociation of molecular hydrogen
through a modification of the chemical potentials of $\rm  H \, \textrm{\scriptsize{I}}$ and $\rm H_2$ (section 2). 
We find that the strong interactions in the dense, helium-rich atmosphere 
results in a significant decrease in the dissociation fraction of molecular hydrogen, with a corresponding change in the 
$\rm H_{2}-He$ Collision-Induced Absorption (CIA) opacity, which is linear function of $n_{\rm H_{2}}$. 
In section 3, we illustrate the impact of the interactions on the $\rm H_{2}/H \,\textrm{\scriptsize{I}}$ ratio on a sequence of white dwarf atmosphere models
with $T_{\rm eff}\rm=4000 \, K$, a gravity of $\rm log \ \it g \rm = 8$ (cgs), and a homogeneous composition of ${\rm He/H} =10^{2},10^{4}, \ \rm and \ 10^{6}$,
where He/H is the number abundance ratio.

\section{The dissociation equilibrium of molecular hydrogen in a dense fluid}
\subsection{Theoretical approach}
The condition for chemical equilibrium (at a given density and temperature) for the dissociation reaction:
\begin{equation} \rm H_{2} \rightleftharpoons 2H \,\textrm{\scriptsize{I}} \label{1}\end{equation}
is given by \citep{CG}
\begin{equation} \mu_{\rm H_2}\rm -2 \it \mu_{\rm H \,\textrm{\tiny{I}}}\rm = 0,\end{equation}
where $\mu_{i}$ is a chemical potential of the species $i$ expressed as
\begin{equation} 
\mu_i= E_{0,i}+k_{B}T\ln\frac{n_{i}h^{3}}{Z_i(2\pi m_i k_B T)^{3/2}} + \mu_i^{non-id}. \label{91} 
\end{equation}
In the above equation, $k_{B}$ is the Boltzmann constant, $h$ the Planck constant, $T$ the temperature, 
$E_{0,i}$ the ground state energy, $n_{i}$ the number density, $Z_i$ the unperturbed internal partition function, and $m_{i}$ is the mass.
The first two terms on the r.h.s. of equation (\ref{91}) represent the ideal contributions of translational and internal degrees of freedom 
and $\mu_{i}^{non-id}$ is the non-ideal contribution to the chemical potential arising from the interparticle interactions in the fluid.
Setting $\mu_{i}^{non-id}\rm =0$ we recover the standard Saha equation for dissociation of molecular hydrogen:
\begin{equation} \beta^{id}=\frac{n_{\rm H_{2}}}{n_{\rm H \,\textrm{\tiny{I}}}^{2}}=\frac{Z_{\rm H_{2}}}{Z^{2}_{\rm H\,\textrm{\tiny{I}}}}\left[\frac{m_{\rm H_{2}}h^{2}}{2\pi m_{\rm H\,\textrm{\tiny{I}}}^{2} k_{B}T }\right]^{3/2}e^{D_{0}/{k_{B}T}} \label{3} \end{equation}
where $D_{0}=2E_{0,\rm H\,\textrm{\tiny{I}}}-E_{0,\rm H_{2}}=\rm 4.478 \, eV$ is the dissociation energy of the isolated hydrogen molecule.
Even for trace species, like $\rm H\,\textrm{\scriptsize{I}}$ or $\rm H_{2}$ in dense helium, the $\mu_{i}^{non-id}$ 
which arise from interactions with the atoms are not negligible and in principle should be comparable in magnitude to $\mu_{He}^{non-id}$.

If we define the quantity $\Delta I$ as
\begin{equation} \Delta I = \mu^{non-id}_{\rm H_{2}}-2\mu^{non-id}_{\rm H\,\textrm{\tiny{I}}}, \label{10}\end{equation}
the non-ideal equilibrium equation can be written in the following form
\begin{eqnarray} \beta = 
\frac{n_{\rm H_{2}}}{n_{\rm H\,\textrm{\tiny{I}}}^{2}}
= \frac{Z_{\rm H_{2}}}{Z^{2}_{\rm H\,\textrm{\tiny{I}}}}\left[\frac{m_{\rm H_{2}}h^{2}}{2\pi 
m_{\rm H\,\textrm{\tiny{I}}}^{2} k_{B}T}\right]^{3/2}e^{(D_{0}-\Delta I)/{k_{B}T}} \nonumber 
\end{eqnarray}
\begin{equation} 
 =  \beta^{id}e^{-\Delta I/k_{B}T}. \label{6}
\end{equation}

Comparing (\ref{3}) with (\ref{6}), we see that the non-ideal effects on the dissociation equilibrium can be \it interpreted \rm as a change in the dissociation energy by a value of $\Delta I$. 
For simplicity, we will follow this interpretation hereafter. We emphasize that this description of the non-ideal contribution to 
the chemical equilibrium (Eq. 6) is {\it identical} to the occupation probability formalism of \citet{HM} if there
is only one bound state in the partition function (e.g. low temperature $\rm H \, \textrm{\scriptsize{I}}$) or if
the $\mu_{i,j}^{non-id}$ is the same for all bound states $j$ of the species $i$
($\mu_{i,j}^{non-id}=\partial f(V,T,{n_{i,j}})/\partial n_{i,j}$ remains constant in Eq. (2.17) of \citet{HM}).
In both cases, the occupation probability can be factored out of the partition function 
and written as $e^{-\Delta I/k_BT}$ (Eq. 6).

In the atmosphere of cool white dwarfs, hydrogen exist mostly as $\rm H_2$ and $\rm H\,\textrm{\scriptsize{I}}$, 
and the $\rm H\,\textrm{\scriptsize{I}}/H_2$ ratio is governed by reaction (\ref{1}) only. For a given temperature, density $\rho$, and the atmosphere composition $y=\rm He/H$, 
the number densities of $\rm H_2$ and $\rm H\,\textrm{\scriptsize{I}}$ are:
\begin{equation} n_{\rm H\,\textrm{\tiny{I}}}=\frac{-1+\sqrt{1+8\beta n_{\rm tot}}}{4\beta} \label{12},\end{equation}
\begin{equation} n_{\rm H_{2}}=\it \beta n^{\rm 2}_{\rm H\,\textrm{\tiny{I}}} \label{13},\end{equation}
where
\begin{equation} n_{\rm tot}=n_{\rm H\,\textrm{\tiny{I}}}+ 2 n_{\rm H_{2}}=\frac{\rho}{m_{\rm H}+{y}m_{\rm He}}. \label{14}\end{equation}
refers to the hydrogen species only. 

\begin{figure*}[t]
\centering
\epsscale{1.75}
\plotone{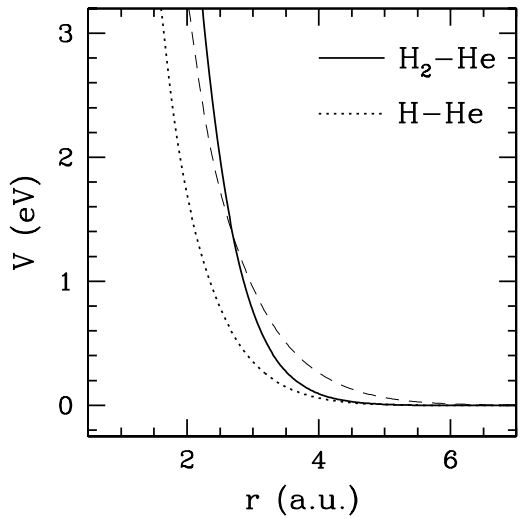} 
\caption{The $\rm H-He$ and $\rm H_2-He$ pair interaction potentials.
The dashed line represents the radius (horizontal axis) of a sphere whose volume equals the effective volume of two H atoms 
colliding with a kinetic energy of ${3\over 2}k_BT$ (see text).}
\end{figure*}

\subsection{Computation of the
$\rm H_2$ dissociation equilibrium}
The non-ideal contributions to the chemical potentials of $\rm H\,\textrm{\scriptsize{I}}$ and $\rm H_2$ were obtained through the numerical solution of the Ornstein-Zernike equation 
in the Percus-Yevick (PY) approximation \citep{M}. For the 
$\rm H - He$ interaction we use the pair potential of \citet{SH}, and for the $\rm H_2 - He$ interaction the pair potential given by \citet{RE} (Figure 1). 
Both potentials are from \it ab initio \rm quantum mechanical calculations and are in good agreement with the independent calculations of \citet{TY} for $\rm H-He$ 
and of \citet{TAO} and \citet{SG} for $\rm H_2-He$. 
As we consider a helium-dominated mixture ($\rm He/H\wig>10^{2}$), the $\rm H-H$, $\rm H_2 - H$, and $\rm H_2 - H_2$ interactions can be neglected.  

High pressure experiments have shown that \it ab initio \rm pair potentials are too repulsive to describe dense systems where N-body effects become important \citep{Nellis,Ross}.
The softening of the pair potentials at high densities can only be quantified experimentally or, alternatively, estimated with N-body quantum mechanical calculations.
Since neither are available for mixtures of trace hydrogen in helium, we resort to \it ab initio \rm potentials. The net effect on the dissociation equilibrium depends
on the \it relative \rm softening of the potentials (Eq. 5) and is therefore less sensitive to N-body effects than the individual potentials.
We also calculated the chemical potentials in 
the Hypernetted Chain approximation and found them to agree within $5\%$ with the PY values up to $\rm 2 \, g/cm^{3}$. 
Since the PY approximation is better suited for short range potentials such as the ones we use here, 
we estimate that our PY calculations are reliable up to at least $\rm 2 \, g/cm^{3}$.

\begin{figure*}[t]
\centering
\epsscale{1.75}
\plotone{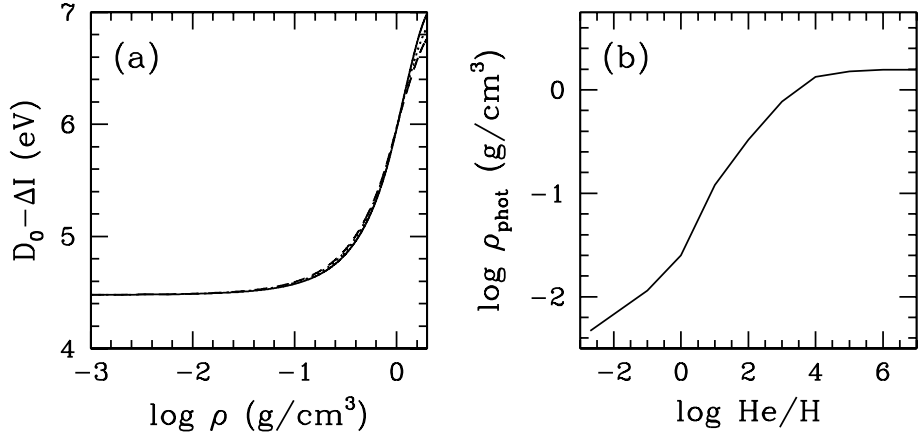} 
\caption{(a) Variation of the effective dissociation 
energy $D_{0}-\Delta I$ as a function of the density of helium for temperatures of
$\rm 3000\, K$ (solid line), $\rm 4000\, K$ (dotted line), and $\rm 5000\, K$ (dashed line). (b) The density $\rho$ at the photosphere 
of cool white dwarf atmosphere models of $T_{\rm eff}\rm=4000\, K$, $\rm log \ \it g \rm = 8$ (cgs), 
and various $\rm He/H$ compositions.}
\end{figure*}

For the internal partition functions $Z_i$, we use expressions for the electronic ground state of the unperturbed hydrogen molecule accounting 
for the vibrational/rotational excitations \citep{HH}, 
and set $Z_{\rm H \,\textrm{\tiny{I}}}=2$ for hydrogen atom. This approximation is justified as the electronic excitation energies of both species are large 
and for temperatures of a few thousands degrees the populations of the electronic excited levels are extremely small. 
However, there is significant thermal excitation of the rotational and vibrational levels of $\rm H_2$ 
and the effect of the dense medium on $Z_{\rm H_2}$ must be considered.  Since the molecule
does not have spherical symmetry, it is possible that its rotation modes will be hindered by
interactions with neighbors at very high density.  Furthermore, the energies of the vibrational
levels, which are spaced by a mere $0.54\,\rm eV$, could also be significantly shifted by these 
interactions. Both of these
effects are discussed by \citet{SCH} and can be neglected under the
present conditions.  The use of the unperturbed rotational and vibrational levels of $\rm H_2$ is thus
justified in the calculation of $Z_{\rm H_2}$.

\begin{deluxetable}{crrrrrrrrrrr}[hb]
\tablecaption{The change in the dissociation energy 
$\Delta I$ (eV) for representative densities and temperatures.}
\tablewidth{0pt}
\tablehead{& \colhead{
$\rho$ $\rm (g/cm^{3})$}   
&& \colhead{$T \rm (K) : \ \ \ \ 2000\rm$} 
& \colhead{$4000\rm$} 
& \colhead{$6000\rm$} & \colhead{$8000\rm$}}
\startdata

 &0.001 &&$-$0.001 &$-$0.001 &$-$0.001 &$-$0.001 \\
 &0.010 &&$-$0.007 &$-$0.010 &$-$0.011 &$-$0.012 \\
 &0.100 &&$-$0.083 &$-$0.108 &$-$0.120 &$-$0.127 \\
 &0.250 &&$-$0.242 &$-$0.297 &$-$0.323 &$-$0.334 \\
 &0.500 &&$-$0.579 &$-$0.668 &$-$0.697 &$-$0.702 \\
 &0.750 &&$-$1.006 &$-$1.075 &$-$1.088 &$-$1.070 \\
 &1.000 &&$-$1.471 &$-$1.494 &$-$1.469 &$-$1.414 \\
 &1.500 &&$-$2.011 &$-$2.094 &$-$1.963 &$-$1.825 \\
 &2.000 &&$-$2.189 &$-$2.387 &$-$2.192 &$-$1.988 \\

\enddata
\end{deluxetable}

\begin{figure*}[t]
\centering
\epsscale{1.75}
\plotone{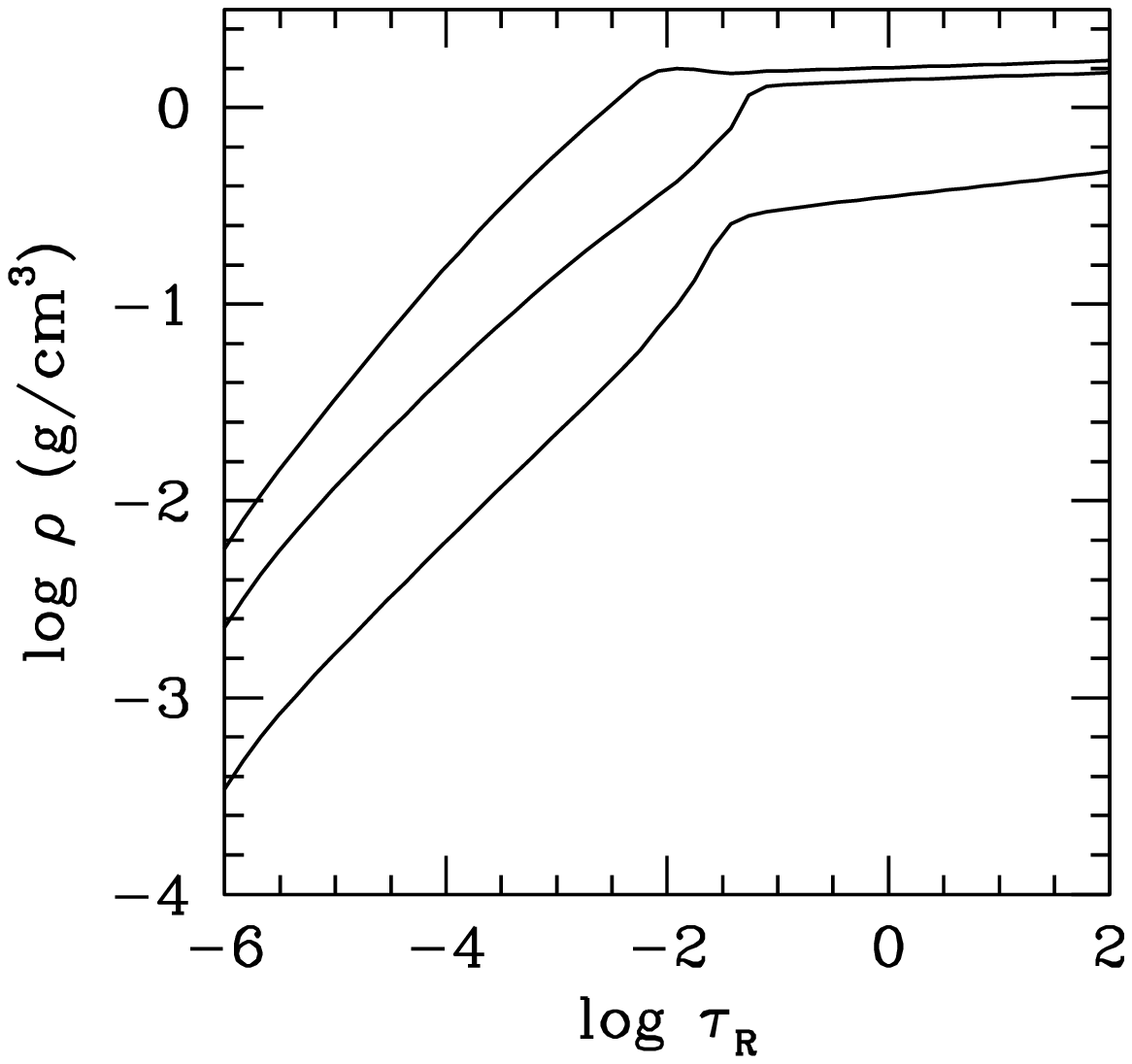} 
\caption{The density profiles of white dwarf atmosphere models of 
$T_{\rm eff}\rm=4000\,K$, $\rm log \ \it g \rm = 8$ (cgs), 
and $\rm He/H$ composition of $\rm He/H=10^6,10^4 \ and \ 10^2$ (from top to bottom, respectively).}
\end{figure*}

More importantly,  the excited rotational and vibrational 
levels of the $\rm H_2$ molecule may be 
differentially affected by the interactions.  This relates to the general problem of the
cutoff of the sum over states
in the internal partition function and is the subject of an extensive 
literature. This truncation of the partition function sum is
often described by an occupation probability formalism \citep{HM} and in the case of interactions 
between neutral particles, is generally described in the
context of the volume occupied by each bound state.  In the case of a diatomic molecule, the volume
occupied increases with the rotational quantum number (centrifugal stretching) and with
the vibrational quantum number (larger amplitude of vibration).  The first effect is very small 
and can be safely neglected.  We estimate the vibrational stretching by considering the change in 
the equilibrium internuclear separation  of the molecule with the vibrational quantum number.
Because of the anharmonicity of the potential, higher vibrational levels correspond to larger
equilibrium separations and a larger average molecular volume.
Based on the potential curve of the $\rm H_2$ molecule \citep{KW} we find that
for $\rm T < 6000\,K$ and densities of up to $\rm 2\,g/cm^3$, the vibrational excitation is
largely limited to the lowest five levels and the partition function is reduced by a few percent
when the larger volume of the excited states is taken into account.
This is a small effect which can be neglected in view of the other uncertainties in the model
such as the $\rm N$-body effects on the interaction potentials.  Thus, the use of the internal partition 
function of the isolated $\rm H_2$ molecule, with all states being affected identically by the interactions
with He (Eq. 6) is a very good approximation.

\section{Results and Discussion}  
\subsection{The H$_2$ dissociation equilibrium in fluid helium}
In general, the $\Delta I$ defined by (\ref{10}) is a function of the density of helium and the temperature (Fig. 2a), but is independent of $\rm He/H$
ratio if $\rm H$ is a trace species ($\rm He/H \wig>10^2$). For the conditions 
in white dwarf atmospheres, the temperature dependence is weak. The change in the dissociation energy $\Delta I$ is negative, making $\rm H_2$ more stable in dense helium.
This may be qualitatively understood by comparing the effective volume occupied by one $\rm H_2$ molecule to that of two $\rm H$ atoms in He. 
Lets assume that these effective volumes are the spaces around each of the particles where the energy of their interaction with $\rm He$ atoms is greater than 
the average thermal kinetic energy of the particles ${3\over 2}k_BT$. This radius is the classical distance of closest approach in a collision.
On Figure 1, we plotted the radius of a sphere whose volume equals 
the effective volume of two hydrogen atoms, as a function of $k_BT$ (dashed curve). The corresponding radius for the hydrogen molecule is represented by the
$\rm H_2-He$ potential (solid curve). 
The effective volume of $\rm H_2$ is smaller than that of two $\rm H$ atoms for $k_BT \wig<1\,$eV.
Since the exclusion of a greater volume results in a decrease of the entropy of He, $\rm H_2$ is more stable in dense helium than two H atoms. 
In dense hydrogen, we have the opposite situation where H$_2$ is less stable at high density (leading to pressure dissociation
of H$_2$) because H$_2$--H$_2$ is more repulsive than the sum of the H--H and H--H$_2$ interactions \citep{SCH}.

\begin{figure*}
\centering
\epsscale{1.75}
\plotone{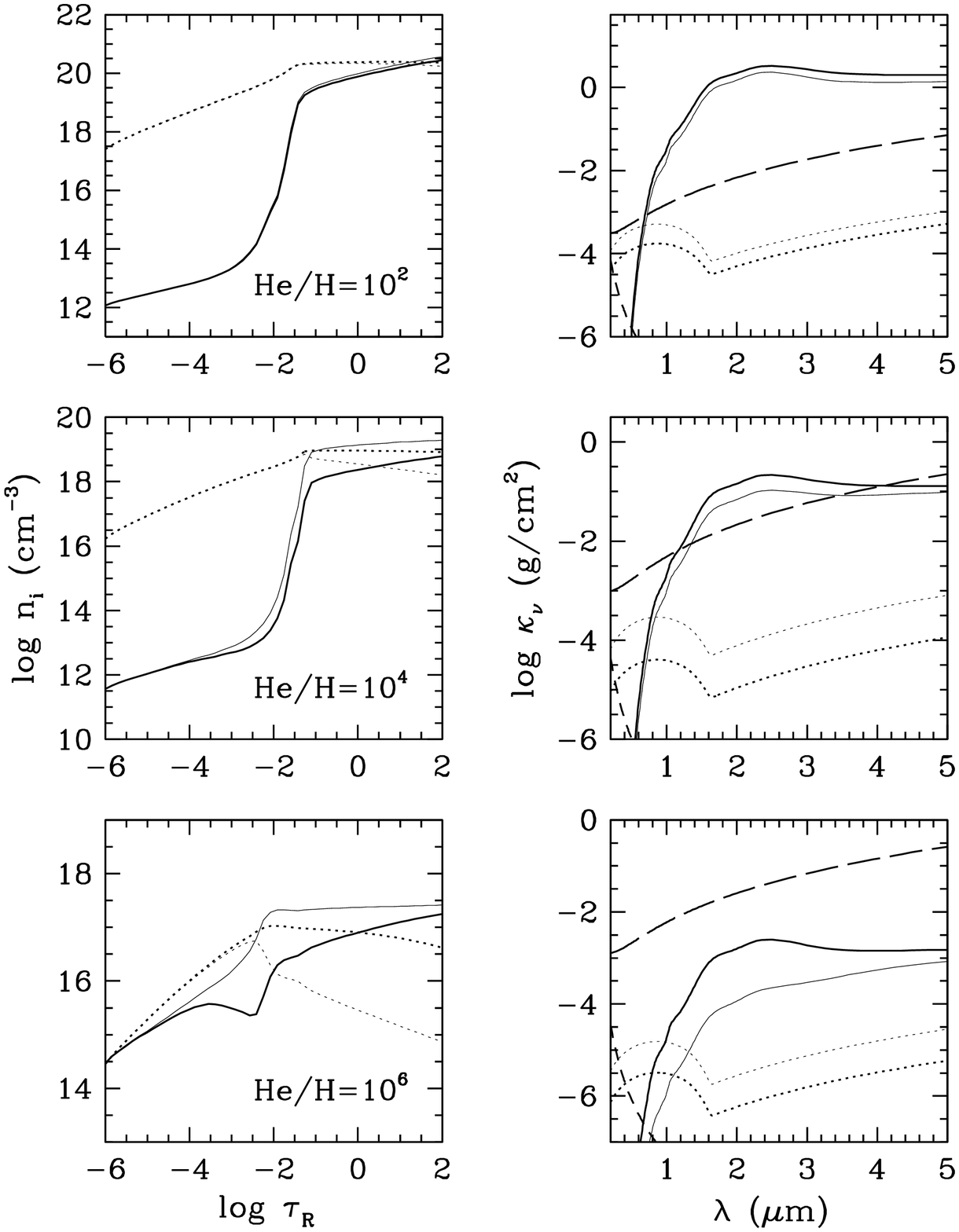} 
\caption{Left panels: The number density of atomic (solid) and molecular (dotted) hydrogen with (thick lines) and without (thin lines)
the non-ideal chemical equilibrium along the atmosphere profiles of Figure 3.
Right panels: Corresponding contributions to the photospheric opacity: $\rm H_2-He$ CIA (solid),  
$\rm He^{-}$ free-free of Iglesias, Rogers, \& Saumon (2002) (long dashed), $\rm H^-$ bound-free 
and free-free (dotted) and Rayleigh scattering (short dashed).}
\end{figure*}

Numerical values for $\Delta I$ obtained from our model are given in Table 1. 
Together with relations (\ref{6})--(\ref{14}) this table allows for an easy implementation of the non-ideal abundances of atomic and molecular hydrogen 
into white dwarf atmosphere codes.

The non-ideal recombination of H becomes significant when $-\Delta I/k_BT\wig> 1$.  For an atmosphere with $T_{\rm eff}=4000\,$K,
this will occur at the photosphere if  $\rho \wig>0.3\,$g/cm$^3$, a density easily achieved in He-rich models. 
As $D_0-\Delta I$ increases with increasing helium density, a lower dissociation fraction will result. 
A strong deviation from the ideal-gas abundance of $\rm H_2$ is expected.   

\subsection{The $\rm H_2$ dissociation equilibrium in He-rich white dwarf atmospheres}
We have applied the non-ideal correction to the $\rm H_2$ dissociation equilibrium to very cool white dwarf atmospheres. 
To show the effect, we solved for the chemistry and opacity along fixed $T-\rho$ white dwarf atmosphere profiles of $T_{\rm eff}=4000\,\rm K$, 
a gravity of $\rm log \ \it g \rm = 8$ (cgs), and various $\rm He/H$
ratios. Figure 3 shows the atmospheric density profiles. These profiles were obtained with a code that solves for static, 
plane-parallel, and LTE white dwarf atmospheres in thermal and hydrostatic equilibrium, accounting for refraction \citep{KS04}. 
The resulting new abundances of $\rm H\,\textrm{\scriptsize{I}}$ and $\rm H_2$ in our models are given on Figure 4 (left panels). As expected, the amount of molecular hydrogen 
increases significantly throughout the atmosphere for $\rm He/H\wig>10^{3}$, and $\rm H_2$ can become the dominant hydrogen species at the photosphere. 

The most important sources of opacity in helium-rich white dwarf atmospheres are $\rm He^{-}$ free-free, Rayleigh scattering, and $\rm H_{2}-He$ CIA 
\citep{HN}. Several physical and chemical effects that alter these opacity sources have already been discussed: the change in the 
free-free opacity and the Rayleigh scattering of helium, as caused by the strong correlations in the dense fluid \citep{IRS}, the change in the number density 
of free electrons \citep{KSM05,BSW}, the presence of heavy elements \citep{Bergeron01}, and the formation of trace species like $\rm He_{2}^{+}$ \citep{ML} and $\rm HeH^{+}$ \citep{HR}.
The effect on non-ideal chemistry on the $\rm H_2$ dissociation on the opacity is shown in the right hand panels of Figure 4. The increase in 
number density of molecular hydrogen results in an increase of 
the $\rm H_2-He$ CIA opacity. This follows from the linear dependence of $\rm H_2-He$ CIA opacity on $n_{\rm H_2}$. 
On the other hand, the $\rm H^-$ bound-free and free-free opacities are reduced, due to the decrease in the abundance of atomic hydrogen.    
Based on the atmospheric structures used here, the effects of the non-ideal chemistry of H$_2$ in dense helium are maximal for He/H $\sim 10^3$.  This
arises from a competition between the need for a high He/H ratio to increase the non-ideal effects by increasing the density at the photosphere (Fig. 2b)
and the need for a high enough hydrogen content in the atmosphere so that H$_2$ (or H) contributes to the total opacity (Fig. 4, right hand side).

\begin{figure*}[t]
\centering
\epsscale{1.75}
\plotone{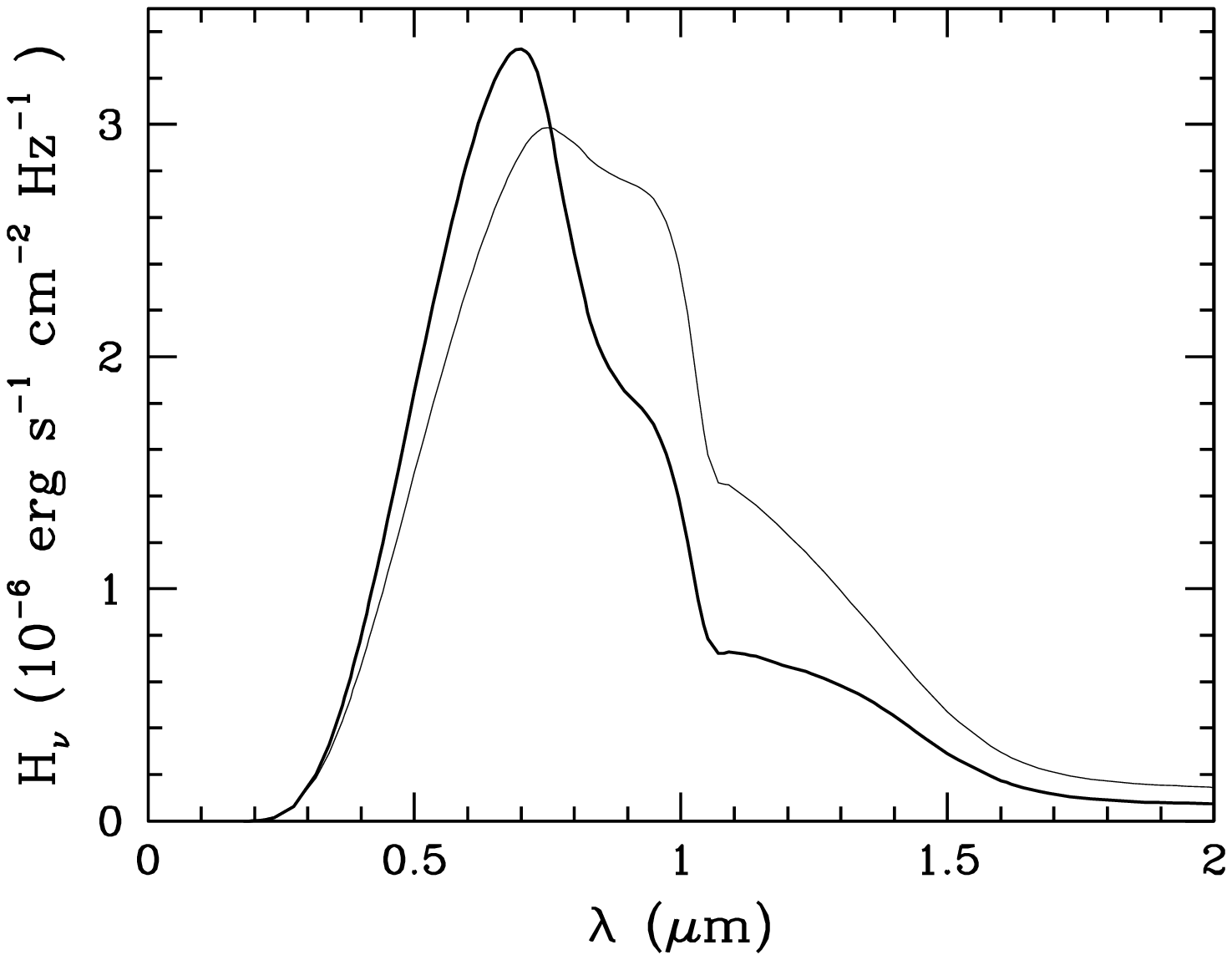} 
\caption{The synthetic spectrum for the white dwarf atmosphere models of 
$T_{\rm eff}\rm=4000\,K$, $\rm log \ \it g \rm = 8$ (cgs), 
and $\rm He/H=10^{3}$ with (thick line) and without (thin line) the non-ideal dissociation equilibrium of $\rm H_2$.}
\end{figure*}

The impact of the non-ideal dissociation equilibrium of $\rm H_{2}$ on the synthetic spectrum of a He-rich white dwarf model is shown in Figure 5.
For a white dwarf atmosphere model of $T_{\rm eff}\rm=4000K$, $\rm log \it \ g\rm=8$, and $\rm He/H=10^{3}$, the rise in abundance 
of $\rm H_2$ increases significantly the opacity in the infrared, causing a redistribution of the flux toward 
shorter wavelengths. This effect on the spectrum of cool white dwarfs is largest 
for $T_{\rm eff}\rm =4000K-4500K$, where $\rm H_2$ is partially dissociated. At lower effective temperatures hydrogen exists mostly 
in molecular form and the effect of the non-ideal dissociation equilibrium 
on the spectrum vanishes at $T_{\rm eff}\rm \sim 3000K$. 

The determination of the $\rm He/H$ composition of very cool He-rich white dwarfs depends mostly on the relative $\rm He-H_2$ CIA, and $\rm He^-$ free-free 
opacities (Fig. 4, right hand side), Since the non-ideal recombination increases the CIA opacity, this effect, taken by itself, will result in a higher 
value for the $\rm He/H$ ratio inferred from a given stellar spectrum. On the other hand, current calculations 
for the ionization fraction of dense helium and the low-frequency behavior of the $\rm He^{-}$ free-free opacity
are uncertain and the $\rm He^{-}$ free-free opacity could be underestimated by as much as $\sim 2-4$ orders of magnitude \citep{KSM05}. It is therefore too early to
draw conclusions about the atmospheric composition of the coolest white dwarfs. 
However, we emphasize that, as the chemical equilibrium (\ref{6}) is not affected by the number of 
free electrons in the atmosphere, our solution for the non-ideal abundance of $\rm H_2$ will not be affected by weak 
ionization of $\rm He$, and the results are limited only by the validity of the $\rm H_2-He$ and $\rm H-He$ interaction potentials.

The non-ideal dissociation equilibrium in pure hydrogen atmospheres is discussed in \citet{SJ}. As hydrogen is much more opaque than helium, 
the density at the photospheres of these stars
is much smaller than for helium-rich atmosphere stars of the same effective temperature. The non-ideal effects are therefore weaker and become important
only in stars of lower effective temperatures and/or higher gravity, where high densities are achieved. \citet{SJ} show that the non-ideality matters in pure-hydrogen atmospheres 
with $\rm T_{eff}\wig<2500K$. White dwarfs with hydrogen-rich atmospheres of such low effective temperatures have not yet been identified.

\section{Conclusions}
Recent discoveries of a number of cool white dwarfs with peculiar spectral energy distribution represent a challenge in the modeling of very cool white dwarf atmospheres, 
as the spectra of these stars cannot be fitted with existing models. We believe that the main reason of this shortcoming of the models is 
the poorly explored, extreme physical regime found inside these atmospheres. We presented a correction to the abundance of $\rm H_2$ that arises 
from the strong interaction of hydrogen molecules and atoms in a dense, fluid helium medium. We have found that in the dense, helium-rich, cool white dwarf 
atmospheres the formation of $\rm H_2$ is more favorable than in the ideal gas description. 
For white dwarfs of $T_{\rm eff}\rm=4000 \, K$ and $\rm He/H\wig>10^{3}$ the abundance of molecular hydrogen increases by an order of magnitude, with a corresponding increase in 
the $\rm H_2-He$ CIA opacity by the same factor. This improvement is a new, significant effect that must be included in realistic modeling 
of very cool white dwarf atmospheres. 

We have shown that the non-ideal effects affect strongly the abundances of even trace species in the atmosphere. This strongly suggest the
necessity of revising the abundances of other trace species with significant opacity for similar effects. We expect that the study of their abundances 
and absorption processes in fluid helium will significantly improve our understanding of the atmospheric physics, composition, and evolution of the oldest and coolest white dwarfs.

I thank D. Saumon for useful discussions and the referee, P. Bergeron, for suggestions that improved the clarity of the manuscript.
This research was supported by the United States Department of Energy under contract W-7405-ENG-36.

\end{document}